\documentclass{article}
\usepackage[a4paper, portrait, margin=1in]{geometry}
\usepackage{graphicx} 
\usepackage[english]{babel}
\usepackage[utf8x]{inputenc}
\usepackage[colorinlistoftodos]{todonotes} 
\usepackage{float}
\usepackage{multirow}
\usepackage[colorlinks, citecolor=cyan]{hyperref}
\usepackage{verbatim}
\usepackage[table]{xcolor}
\usepackage{amsmath}
\AtBeginEnvironment{gather}{
\setlength{\abovedisplayskip}{0pt}
}
\usepackage{adjustbox}
\usepackage[backend=biber,sorting=none, minbibnames=3, style=numeric-comp]{biblatex}
\usepackage{amsmath}
\DeclareMathOperator{\sinc}{sinc}
\usepackage{lineno}

\usepackage[font=footnotesize]{caption}
\usepackage{subcaption}
\usepackage{titlesec}
\titlespacing\section{0pt}{12pt plus 4pt minus 2pt}{0pt plus 2pt minus 2pt}
\titlespacing\subsection{12pt}{12pt plus 4pt minus 2pt}{0pt plus 2pt minus 2pt}
\titlespacing\subsubsection{12pt}{12pt plus 4pt minus 2pt}{0pt plus 2pt minus 2pt}
\titleformat{\section}{\normalfont\fontsize{12}{15}\bfseries}{\thesection.}{1em}{}
\titleformat{\subsection}{\normalfont\fontsize{12}{15}\bfseries}{\thesubsection.}{1em}{}
\titleformat{\subsubsection}{\normalfont\fontsize{12}{15}\bfseries}{\thesubsubsection.}{1em}{}
\usepackage{enumitem}

\title{\textbf{\huge Analyzer-less X-ray Interferometry with Super-Resolution Methods}}

\usepackage{authblk}

\author[1]{Murtuza S. Taqi}
\author[1]{Hunter C. Meyer}
\author[1*]{Joyoni Dey}
\affil[1]{Department of Physics and Astronomy, Louisiana State University, Baton Rouge, LA, 70803}
\affil[*]{Corresponding Author: deyj@lsu.edu}

\date{} 

\begin{document}

\maketitle

\begin{abstract}

X-ray interferometry provides valuable information in terms of attenuation, small-angle scatter, and differential-phase contrast. This multi-modal contrast can aid in many clinical applications, such as lung
diseases and breast cancer. However, standard interferometry has an analyzer grating that can increase the dose requirement to maintain the same image quality as a standard X-ray.
We propose the use of super-resolution methods for X-ray grating interferometry without an analyzer, with detectors that fail to meet the Nyquist sampling rate needed for traditional image recovery algorithms. Detector phase steps are used to nominally recover the fringe sampling, followed by iterative recovery of the visibility and object parameters. This method enables Talbot--Lau interferometry without the X-ray absorbing analyzer. Removing the absorbing analyzer grating may improve dose efficiency and reduce system complexity. We demonstrate the use of super-resolution methods to iteratively reconstruct attenuation, differential-phase, and dark-field images using simulations of two-dimensional lung phantoms with lesions. A direct CdTe detector was simulated with pixel sizes of 55, 75, and 150 $\mu$m. The simulation results show that the proposed super-resolution iterative reconstruction method for Talbot-Lau Interferometry remains stable under the simulated noise conditions and can recover image parameters in cases where traditional algorithms cannot be used.

\end{abstract}

\section{Introduction}
\label{sec:introduction}

X-ray grating interferometry is a phase-sensitive imaging technique that simultaneously captures attenuation, differential-phase, and dark-field images.  The attenuation image shows X-ray absorption, similar to a traditional radiograph, while the differential-phase and dark-field images are related to X-ray refraction and small-angle scatter. With the introduction of differential-phase and dark-field images, X-ray interferometry has a wide variety of potential applications, including lung imaging \cite{bib:Gassert, bib:Bech, bib:Yaroshenko, bib:Velroyen}, breast imaging \cite{bib:Wang2014, bib:Tapfer, bib:Scherer, bib:Koehler}, arthritis imaging \cite{bib:Stutman, bib:Tanaka}, osteoporosis imaging \cite{bib:Gassert2023}, additive manufacturing quality assurance \cite{bib:Zhao, bib:Brooks} and porosimetry \cite{bib:Revol}.

Various types of grating interferometers exist: the Dual-phase Grating Interferometer (DPGI), Modulated Phase Grating Interferometer (MPGI), and the most prevalent for biomedical imaging applications, the Talbot--Lau Interferometer (TLI). The TLI is at the forefront of preclinical and clinical studies for lung and breast imaging \cite{bib:Gassert, bib:Bech, bib:Yaroshenko, bib:Velroyen,bib:Wang2014, bib:Tapfer, bib:Scherer, bib:Koehler,bib:Stutman, bib:Tanaka}. The traditional Talbot--Lau Interferometer (TLI) \cite{bib:Momose2003, bib:Momose2005, bib:Pfeiffer2006, bib:Pfeiffer2009}, uses a binary phase grating (labeled G1) to produce a high-resolution fringe pattern that cannot be resolved by detectors with clinically relevant pixel sizes. A second grating (labeled G2) serves as an analyzer grating to resolve the sub-pixel fringe patterns. However, since the analyzer is an absorption grating, decreased visibility may reduce the dose by a factor of 5 while maintaining the same CNR \cite{bib:BertilsonPMB2024}. Fabrication of high-aspect-ratio absorbing gratings also remains a technical challenge in X-ray grating interferometry \cite{bib:jefimovs2021fabrication}. Removal of the analyzer grating also reduces cost, simplifies setup and alignment, and can reduce fabrication complexity. 

There are existing solutions to implement an analyzer-less interferometer. One such approach is by Cartier \textit{et al.} using smaller detector pixels of $25 ~\mu m$ with charge sharing analysis to enable a resolution of $1~\mu m$ \cite{bib:cartier2016micrometer}. A similar approach has been adopted by Bertilson \textit{et al.} using a pixel size of $2.5 ~\mu m$ using a virtual grating \cite{bib:BertilsonPMB2024}. These approaches are limited by the small size of the detector pixel. An analyzer-free approach using a Modulated Phase Grating has been employed by our group \cite{bib:JXuHamDey,bib:MeyerDeySciRep2024}. The MPGI requires large unaliased fringe periods to be resolved by the clinically relevant detectors, which can lead to a lower auto-correlation length ($ACL= \lambda \frac{D_{OD}}{P_d}$). Another analyzer-free approach, proposed by Rutishauser \textit{et al.}, uses a structured scintillator that replaces the conventional analyzer grating and functions as both the analyzer element and X-ray-to-light converter \cite{bib:rutishauser2011structured}. This approach is attractive for hard X-ray TLI because it can improve fringe visibility at higher photon energies. However, the scintillator structure must be matched to a specific grating pitch and system geometry, limiting flexibility. In addition, Rutishauser \textit{et al.} reported substantially longer exposure times to achieve comparable detector counts, indicating a potential dose-efficiency tradeoff for clinical applications.

In this work, we extend analyzer-less grating-based interferometry to clinically relevant detector pixel sizes by incorporating super-resolution reconstruction methods \cite{bib:SuperResFarsiu}. The term super-resolution refers to the process of sub-pixel sampling, achieved by incrementally stepping the detector and subsequently applying iterative reconstruction methods to recover the image parameters. This work generally reflects the strategy presented in Gutman \textit{et al.}\cite{bib:SuperResGutman} and other super-resolution applications, which combines multiple low-resolution (LR) images to reconstruct a high-resolution (HR) image. However, our approach necessarily differs from that of Gutman \textit{et al.} work in that a central component is our \textit{imaging} framework, which is pixel localization of attenuation, dark-field and differential-phase. Super-resolution methods have been used by scientists/engineers for decades for various applications \cite{bib:SuperResFarsiu,bib:SuperResDreierXray,bib:SuperResGutman}, including angular super-resolution with standard X-ray systems to measure small-angle X-ray scattering (SAXS) data \cite{bib:SuperResGutman}. Notably, the work in Gutman \textit{et al.} \cite{bib:SuperResGutman} provides the angular distribution of SAXS, without the spatial localization information afforded by grating-based interferometric dark-field imaging, the focus of our present work.

In standard interferometry recovery, where the projected fringe period, $P_d$, is adequately sampled through the analyzer, one of the gratings is typically stepped without the object (reference) and then with the object \cite{bib:Weitkamp2006tomography,bib:Pfeiffer2006}. Two phase-stepping curves are obtained at each pixel, one with and without the object. From the two curves, the attenuation, dark-field (related to small-angle X-ray scatter), and the differential-phase of the object are estimated \cite{bib:Marathe}. 

In our method, the phase-stepped images, obtained by stepping the detector (comprising the low-resolution (LR) images), are interlaced.  Explicit forward model and iterative reconstruction are employed to recover the attenuation, differential-phase, and dark-field images for lung imaging applications. When the required fringe pattern is undersampled by the detector, interlacing the phase-stepped images allows us first to nominally recover a higher sampling for the reference and with-object signals. Then the visibility loss is retrieved for the reference signal, via a forward model of the full system, which in turn is used to compare with the with-object signal iteratively to estimate the object attenuation, dark-field, and differential-phase.

For the Talbot--Lau system, super-resolution enables imaging without the analyzer grating and may improve dose efficiency. Without the analyzer, the fringe period at the detector, $P_d$, must be resolved, depending on the pitch of the $G_1$ grating and its magnification under the cone-beam geometry. The super-resolution method allows this $P_d$ to be smaller than the detector pixel size. Consequently, the analyzer-less method becomes feasible for many cases where fringes would otherwise not be directly resolvable. In this work, we calculate several potential analyzer-less TLI geometries for lung imaging, in which $P_d$ and object parameters can be recovered using a super-resolution method. Eliminating the analyzer grating also reduces system cost and simplifies experimental alignment.

The novelty of this work is the use of detector-stepped super-resolution methods to recover sub-pixel fringe information in analyzer-free grating interferometry. To our knowledge, this approach has not previously been applied to recover attenuation, differential-phase, and dark-field images in an analyzer-free Talbot-Lau configuration with clinically relevant detector pixel sizes. The interlacing procedure provides an effective sampling pitch set by the detector step size rather than the native detector pixel size. However, the final resolution remains limited by noise and residual blurs due to reconstruction inaccuracies.




The goal of this work is to demonstrate, in simulation, the recovery of the \textit{projection images} for the three imaging modalities from detector-stepped images, under-sampled acquisition conditions.


\begin{figure}[htbp]
    \centering
    \includegraphics[width=0.75\textwidth]{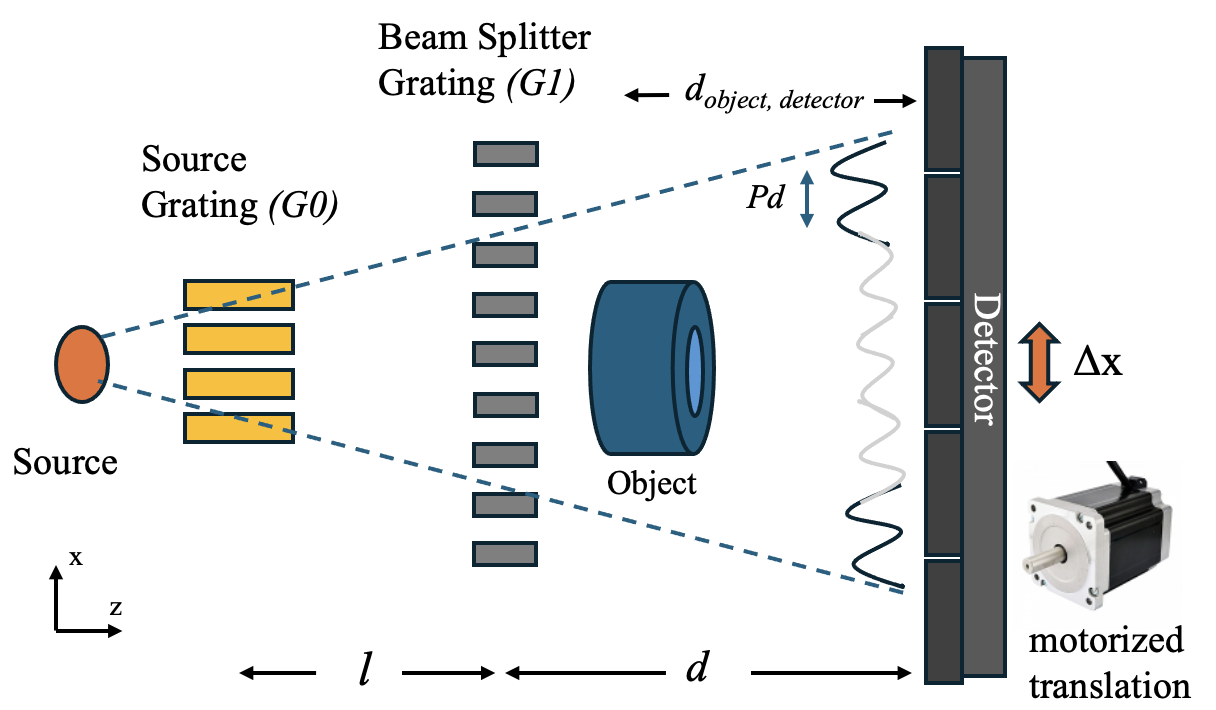}
    \caption{Schematic illustration of the proposed  Analyzer-less Talbot--Lau setup using a motorized translation of the detector.}
    \label{fig:TLI_setup}
\end{figure}

\section{Interferometry Image Generation}
\label{subsec:image_simulation}

\subsection{\textnormal{\textit{Forward-Model-Based X-ray Interferometric Simulation}}}

Two-dimensional interferometry signals are simulated at sub-pixel phase steps by calculating the interference fringes with and without the object in place, referred to as the `reference' and `object' signals.  The simulations are based on the Talbot--Lau system without an analyzer, and the signals are approximated as perfect sinusoids. The source and detector blur are then applied, detector sub-sampling is simulated, and Poisson noise is added.  The reference image for each phase step, $g_r^k(x,y)$, is calculated using Eqn. \ref{eq:reference_signal}, $B(x,y)$ is the per-pixel bias, $t_k(x,y)$ is the per-pixel translation of the k-th phase step, and $P_d$ is the fringe period at the detector.  

\begin{figure}[h!]
    \centering
    \includegraphics[keepaspectratio=true, width=\textwidth, height = 0.7\textwidth]{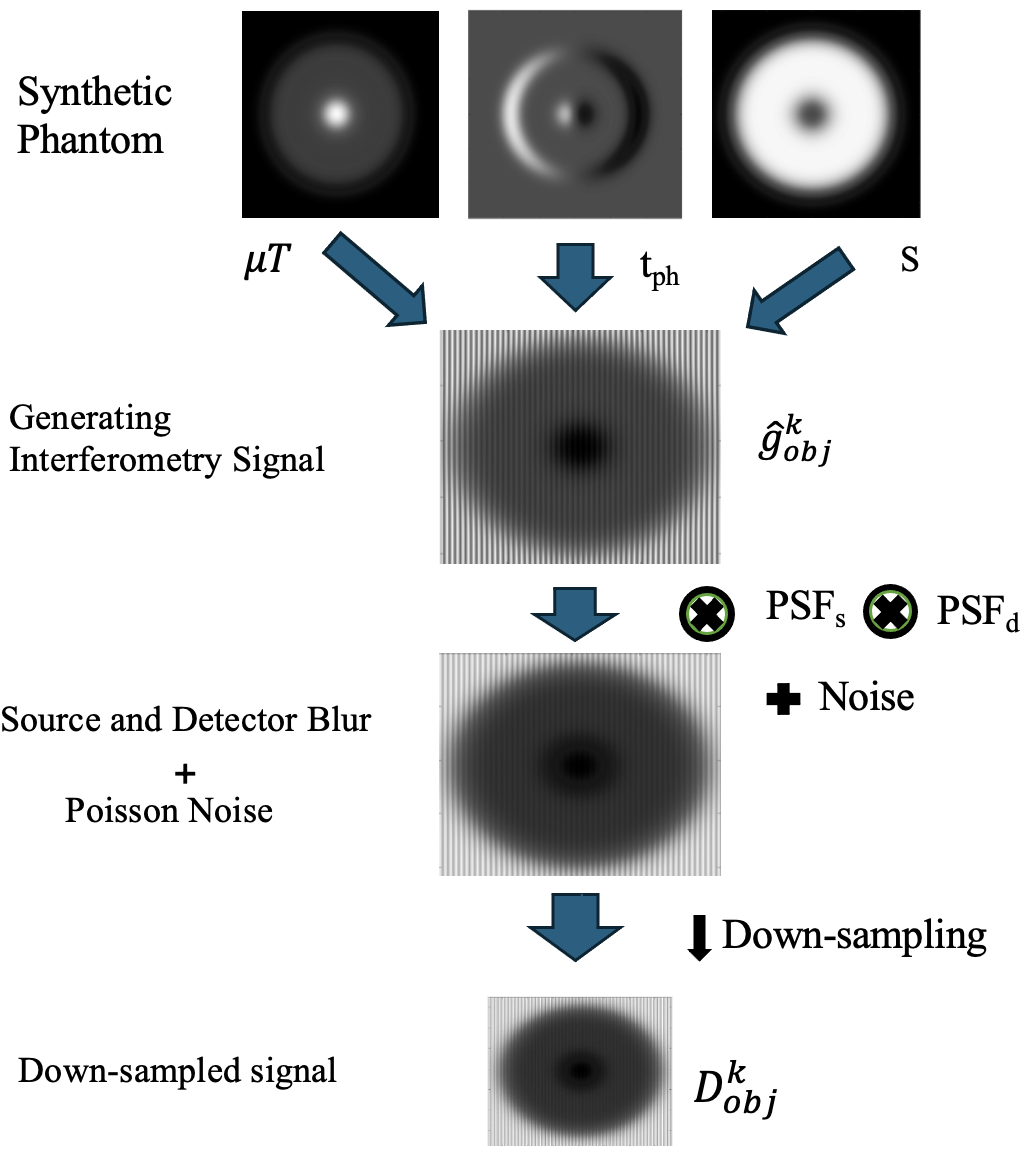}
    \caption{Flowchart of the Phantom generation with known ground truth of $\mu T$, dark field parameter, $S$ and differential-phase-shift($t_{ph}$). The interferometry signal is generated by using equation \ref{eq:object_signal} by using the above-defined parameters. The phase step is obtained by translation of the interferometry pattern, but for illusion purposes, only a single phase step is shown.}
    \label{fig:phantom generation flowchart}
\end{figure}

\begin{equation}
    \label{eq:reference_signal}
    g^k_r(x,y) = \int_{-\infty}^{\infty} \Phi(E) R(E)  B(x,y) \left[ 1+ V(E) \cos \left[ \frac{2\pi}{P_d} \left( x + t_k \right) \right] \right] dE
\end{equation}

Note that the reference $g^k_r(x,y)$ is a 2D image, but the grating is only in one direction. 
The object image is simulated similarly, as shown in Eqn. \ref{eq:object_signal}, where $e^{-\mu T(x,y)}$ represents the object attenuation, $t_{ph}(x,y)$ represents the phase shift due to refraction, and $e^{-S(x,y)}$ represents the visibility loss due to small angle scatter.  The phase shift due to refraction, $t_{ph}(x,y)$, is represented as $t_{ph}(x,y) = \frac{\lambda}{2\pi}\frac{\partial \varphi(x,y)}{\partial x}D_{OD}$ where $D_{OD}$ is the object-to-detector distance and $\frac{ \partial\varphi(x,y)}{\partial x}$ is the partial derivative of the phase. Unlike traditional interferometry, we phase step the \textit{detector}, not the grating, to achieve sub-pixel resolution of the object. Hence, each term is also shifted by $t_k$, since it is the detector that is shifted for phase stepping, not the grating.

\begin{equation}
    \label{eq:object_signal}
    g^k_{obj}(x,y) = \int_{-\infty}^{\infty}  \Phi(E) R(E)  B(x,y) \left\{ 1+ V(E)e^{-S^{t_k}(E)} \cos \left[ \frac{2\pi}{P_d} \left(x + t_{ph}^{t_k}(E) + t_k \right) \right] \right\} e^{-\mu T^{t_k}(E)} dE
\end{equation}

To simulate phase-stepping of the detector, the generated signals $g^k_r(x,y)$ and $g^k_{obj}(x,y)$ are shifted by the step resolution $t_k$, which is typically a few microns. Here, the superscript $t_k$ indicates that the object maps are evaluated at the detector-shifted sampling coordinate for the $k$-th detector step.

\textit{Detector Stepping}: Detector stepping was simulated using translation step sizes of 5 and 10$~\mu$m. These values were selected to represent experimentally achievable micron-scale positioning. Bloomer \textit{et al.} reported a detector translation step size of 3$~\mu$m with a positioning accuracy of $\pm 1~\mu$m, demonstrating that our stepping parameters are feasible \cite{bib:bloomer2020Diamond_xray_detector}. In addition, motorized translation stages used in micro-X-ray imaging systems can provide 1$~\mu$m full-step resolution, with finer effective positioning achievable through microstepping, further supporting the feasibility of the step sizes considered in this study \cite{bib:ruf2020Flexible_micro_x_ray}.

\subsection{\textnormal{\textit{Source and Detector Spectral Considerations}}}

The source spectrum was generated for a tungsten anode operated at 120 kVp with 7 mm of aluminum filtration to reduce beam-hardening effects. The SPEKTR 3.0 software package was used to compute the filtered source spectrum, including the effect of aluminum attenuation \cite{bib:Spektr_punnoose2016}. Additional beam filtration by the $G_0$ grating was incorporated by modeling transmission through a gold $G_0$ grating with a duty cycle of 0.5. The resulting effective spectrum was normalized such that the energy-dependent weighting factors, $w(E)$, satisfied $\int_E w(E),dE = 1 $.

The detector was modeled as a photon-counting detector with a 300$~\mu$m cadmium telluride (CdTe) sensor layer. Detector pixel sizes of 55, 75, and 150 $\mu$m were simulated as representative direct-detector sampling conditions. The 55 and 150 $\mu$m pixel sizes correspond to MiniPix detector pixel size, while the 75 $\mu$m pixel size corresponds to the Eiger detector.
 The detector efficiency, $\eta_{det.}(E)$, was incorporated based on the energy-dependent absorption of the CdTe sensor layer. The detector response function, $R(E)$, was modeled following Schlomka \textit{et al.} \cite{bib:schlomka2008experimental} to account for spectral distortion effects in CdTe detectors, including charge sharing and K-escape. After combining the source spectrum, $G_0$ transmission, detector efficiency, and detector response, the effective spectrum was normalized to its maximum value for visualization. The resulting spectrum is shown in blue in Fig.~\ref{fig:Visibility Curve}, with the detector thresholds indicated in black.

\subsection{\textnormal{\textit{Phantom Design and Material}}}

The digital phantom consisted of a cylindrical soft-tissue tumor insert embedded within inflated lung tissue, with the surrounding region assigned as air. The energy-dependent linear attenuation coefficient, $\mu(E)$, and refractive index parameter, $\delta(E)$, (required for differential phase) were obtained using the XrayDB library, based on tabulated data from Chantler \cite{bib:chantler1995theoretical}. The elemental compositions of the phantom materials were defined according to ICRU Report 44 using the NIST X-ray mass attenuation coefficient database \cite{bib:NIST_XrayMassCoef_Table2}. The energy-dependent linear diffusion coefficient (required to set up small-angle-scatter visibility loss) for healthy lung tissue was modeled using the power-law relationship described by Taphorn \textit{et al.} \cite{bib:taphorn2021directDifferentiate}. The empirical power-law coefficient from  Taphorn \textit{et al.} captures the energy dependence for a fixed interferometer sensitivity, but does not explicitly model the change in dark-field response caused by changing the projected fringe period($P_d$) or sample-to-detector distance. 


The intensities were generated separately for each energy using the corresponding phantom parameters. The energy-specific intensities were then weighted by the spectrum and summed to obtain the spectrum-weighted signal as shown in Eqn \ref{eq:object_signal}. 

The simulated field of view was $15 \times 15~\mathrm{mm}^2$. The lung region had a diameter of 9 mm, and the tumor insert had a diameter of 6 mm. The total phantom thickness along the beam direction was 10 cm, with a tumor thickness of 5 cm. To be able to compare across different detectors, the object was assumed to be the same size at the detector for the different geometries simulated. This means that the actual object was of a different size depending on the magnification afforded by the geometry.
The representative geometries used for the detector cases shown in Fig.~\ref{fig:comparison_grid} are listed in Table~\ref{tab:det cases parameters}. Additional projected fringe periods were simulated to evaluate the dependence of error in recovery of $S$ on the Nyquist undersampling ratio, as shown in Fig.~\ref{fig:error plot}. 

\subsection{\textnormal{\textit{Ground Truth Phantom}}}

To evaluate super-resolution recovery, a ground-truth phantom is generated. The attenuation, phase-shift, and dark-field maps are generated from the energy-dependent material properties obtained as described in the previous section. The attenuation projection, $\mu T$, was calculated from the material energy-dependent linear attenuation coefficients and projected thicknesses. The phase-shift map, $t_{\mathrm{ph}}$, was calculated from the projected phase gradient using $t_{ph}=(\lambda/2\pi)(\partial \varphi/\partial x)D_{OD}$ where the energy dependence enters through $\lambda(E)$ and the refractive index decrement, $\delta(E)$. The final polychromatic ground-truth maps were obtained using the same effective spectral weighting used in the signal generation. The polychromatic dark-field visibility ratio, (D), was calculated from the ratio of the spectrum-weighted attenuated visibility signal to the spectrum-weighted attenuation signal, following the polychromatic dark-field formulation described by De Marco \textit{et al.}~\cite{bib:deMarco2023_XDF_GT}. The projected dark-field signal was then defined as ($S=-\ln(D)$). The corresponding ground-truth maps are shown in the leftmost column of Fig.~\ref{fig:comparison_grid}.

\subsection{\textnormal{\textit{Fringe Visibility}}}
\begin{figure}[htbp]
    \centering
    \includegraphics[width=0.75\textwidth]{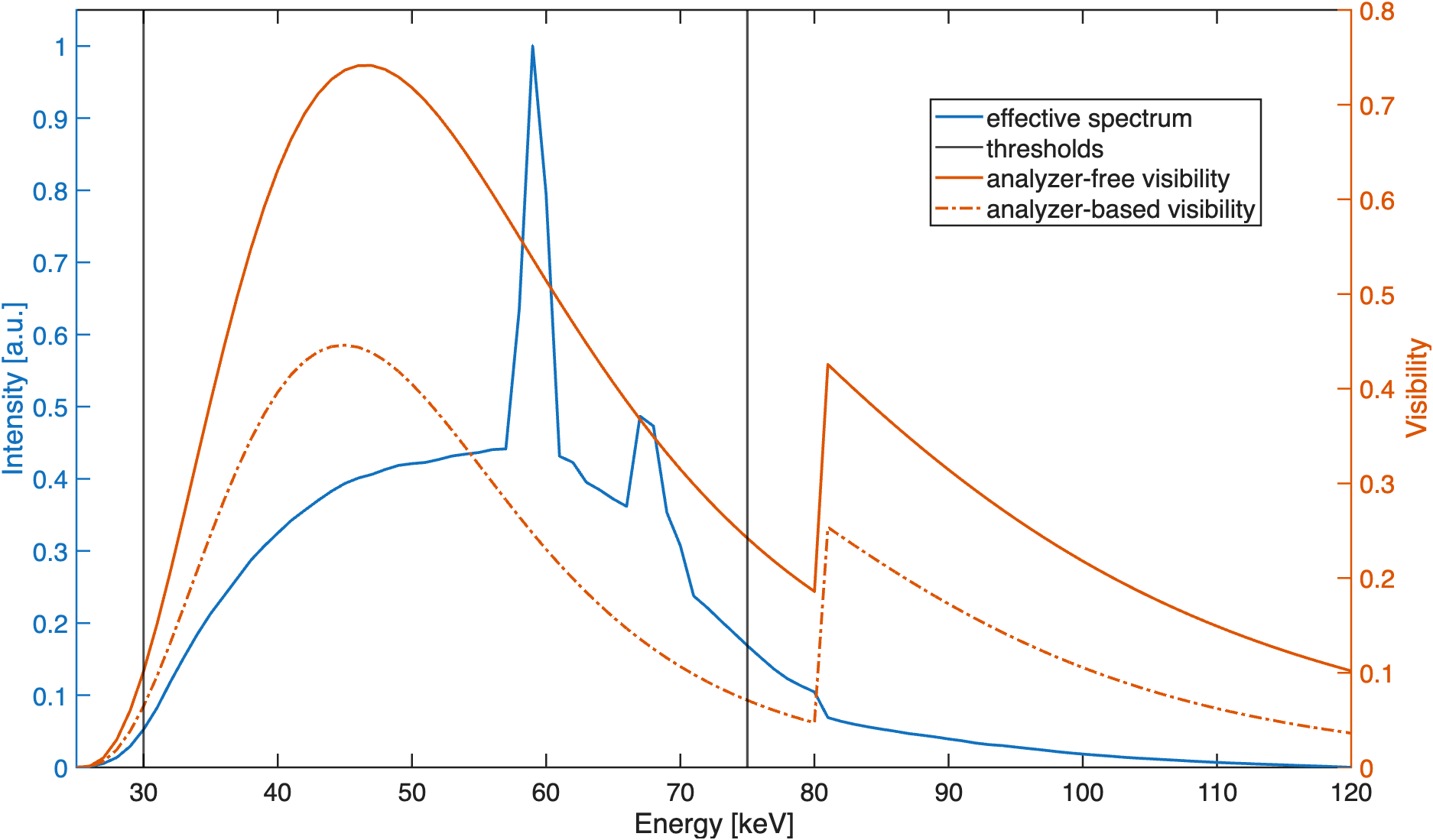}
    \caption{Effective spectrum of the thresholded CdTe detector shown in blue and visibility spectra for the simulated Talbot--Lau interferometer, simulated using visibility equations. The analyzer-based visibility is shown as a red dashed line, while the analyzer-free visibility is shown as a red solid line. Black vertical lines indicate the threshold positions.}
    \label{fig:Visibility Curve}
\end{figure}

The energy-dependent fringe visibility was calculated using the expression derived by Thüering and Stampanoni, with higher-order Fourier coefficients neglected \cite{bib:thuering2014performance}. The transmissions of the $G_0$ and $G_2$ gratings were also incorporated into the visibility model. The analyzer-free visibility was used for the simulations; however, the corresponding analyzer-based visibility curve is also shown in Fig.~\ref{fig:Visibility Curve} for comparison. The analyzer-based visibility curve is consistent with previously reported trends in the literature \cite{bib:mechlem2019spectral,bib:taphorn2021directDifferentiate}. The grating specifications used in the calculation are summarized in Table~\ref{tab:simulation_parameters}.

\begin{equation}
V_{\mathrm{AB}}(E)
=\frac{2a_0\left[1-b_0(E)\right]}{b_0(E)+a_0\left[1-b_0(E)\right]}\sinc_\pi(a_0)\,\frac{a_2\left[1-b_2(E)\right]}{b_2(E)+a_2\left[1-b_2(E)\right]}\sinc_\pi(a_2)\,\frac{2}{\pi}\left|\sin^\eta\left(\frac{\pi E_0}{2E}\right)\sin\left(\frac{m\pi E_0}{2E}\right)\right|.
\end{equation}

\begin{equation}
V_{\mathrm{AF}}(E)
= \frac{2a_0\left[1-b_0(E)\right]}{b_0(E)+a_0\left[1-b_0(E)\right]}\sinc_\pi(a_0)\frac{2}{\pi}\left|\sin^\eta\left(
\frac{\pi E_0}{2E}\right)\sin\left(\frac{m\pi E_0}{2E}\right)\right|.
\end{equation}

where $a_0$ and $b_0$ denote the duty cycle and bar transmission of the $G_0$ grating, respectively, and $a_2$ and $b_2$ denote the corresponding duty cycle and bar transmission of the $G_2$ grating. Note that the bar-transmission is the energy-dependent transmission factor of the grating bars. The parameter $m$ represents the Talbot order, and $\eta$ defines the phase shift of the phase grating. A $\pi$ phase grating was used in this simulation, corresponding to $\eta = 2$. The $V_{AB}$ shown in Fig. \ref{fig:Visibility Curve} uses the same grating parameters for $G_2$ as the $G_0$. The energy-dependent grating bar transmission was calculated using attenuation data from the NIST X-ray mass attenuation coefficient database.

\subsection{\textnormal{\textit{Source and Detector Blur}}}

Source blur was modeled by representing the finite focal spot as a rectangular source distribution. The projected blur at the detector was scaled by the source magnification factor, $M_s = \frac{d}{l}$,
where $d$ is the grating-to-detector distance and $l$ is the source-to-grating distance. The corresponding source point-spread function, $PSF_s$, was modeled as a box kernel whose effective width was determined by the G0 aperture projected onto the detector by pinhole projection through G1. This $G_0$-masked source blur reduces the fringe modulation and therefore directly affects the simulated fringe visibility. As for the focal spot size of the source, a 0.5 mm FWHM source size was used in the simulation.

The detector response was modeled by incorporating a presampling point-spread function, $PSF_d$, for the direct photon-counting detector. The detector $PSF_d$ accounts for charge-sharing and threshold effects and was based on the charge-sharing and noise distributions described by Zambon \textit{et al.} \cite{bib:zambon2018fitting} and Stierstorfer \textit{et al.} \cite{bib:stierstorfer2019modeling}. Charge sharing was parameterized using a charge-cloud size of $\sigma_c = 15~\mu\mathrm{m}$ rms, while electronic noise was represented by $\sigma_n = 1.5~\mathrm{keV}$ rms. The polychromatic dependence of $PSF_d$ was not incorporated; instead, $PSF_d$ was generated at a representative mean energy of 70 keV with a threshold energy of 30 keV. Because the effective charge-sharing and noise behavior may depend on detector pixel size and thresholding conditions, conservative values of $\sigma_c$ and $\sigma_n$ were selected for the simulation. The detector $PSF_d$ was assumed to be shift invariant.

\textit{Noise:} The Poisson noise was simulated at levels determined by analyzing the signal-to-noise ratio (SNR) in previous experiments \cite{bib:MeyerDeySciRep2024}. The mean intensity could not be used directly due to calibration factors. Therefore, we focused on the SNR = mean/std = $\sqrt{N}$ to arrive at the effective bias count, N. The measured SNR was approximately 200 for each phase step image. Assuming Poisson noise, this indicates about 40K counts (pre-calibration). The images were taken under low-dose operation, with a Microfocus X-ray tube running at a $55~\mu A$ current and $40~kVp$ voltage for 20-second exposures. 
The bias term, $B$, was not modeled explicitly as an energy-dependent quantity. Instead, it was approximated by a spectrum-weighted mean value over the effective source spectrum. This approximation reduces the complexity of the polychromatic model while retaining the average contribution of the baseline signal. A similar energy dependence of the baseline term has been reported by Wilde and Hesselink, who showed that the average baseline value, $B_0(E_i)$, increases with energy for a gold grating as the gold attenuation decreases \cite{bib:wilde2020modeling}. The bias was adjusted for different phase steps to keep the total counts the same as the aforementioned case. To keep total counts over all phase steps about the same as the experimental case (N=25), the simulations were scaled by the number of phase steps. For N=15,  bias counts  = $ 67K$ for each step were used, and for N = 5, bias counts = $200K$ for each step.

\begin{gather}
    \label{eq:blurred_reference_signal}
    D^k_r(x,y) = [g^k_r(x,y) \star \star PSF_s(x,y) \star \star PSF_d(x,y) + noise(x,y)]_{\downarrow} \\
    \label{eq:blurred_object_signal}
    D^k_{obj}(x,y) = [g^k_{obj}(x,y) \star \star PSF_s(x,y) \star \star PSF_d(x,y)+noise(x,y)]_{\downarrow}  
\end{gather}

Poisson noise is added to the N phase step signals and down-sampled (indicated by the ${\downarrow}$) to the detector pixel rate to get  $D^k_r(x,y)$ and $D^k_{obj}(x,y)$. They are then used as inputs into the image recovery algorithm for the iterative reconstruction of $\mu T(x,y)$, $S(x,y)$, and $t_{ph}(x,y)$.

\begin{table}[htbp]
\centering
\caption{Values for the parameters used in the simulation of the phantoms.}
\label{tab:simulation_parameters}

\renewcommand{\arraystretch}{1.15}
\begin{tabular}{@{}l c r@{}}
\hline
Parameter & References & Value \\
\hline
Magnified $G_0$ blur  & s & $10, 15, 20~\mu$m \\
$G_0$ duty cycle & $a_0$ & 0.5 \\
$G_1$ period & $p_1$ & $7.75~\mu$m \\
$G_1$ duty cycle & $a_1$ & 0.5 \\
Grating design energy & $E_{0}$ & 50 keV \\
No. of detector steps & $N$ & 15, 11 \\
Step resolution & $t_k$ &  5, 10 $\mu m$ \\
Talbot order & $m$ & 1 \\
Grating phase parameter & $\eta$ & 2 \\
$G_0$ thickness &  & Gold, $200~\mu$m \\
$G_1$ thickness &  & Silicon, $64 ~\mu$m \\
\hline

\end{tabular}
\label{tab:simulation parameters}
\end{table}

\begin{table}[htbp]
\centering
\caption{Representative geometry and acquisition parameters for the simulated analyzer-less Talbot--Lau detector cases with matched sensitivity.}

\label{tab:geometry_cases}
\begin{tabular}{|c|c|c|c|c|c|c|}
\hline
\rowcolor{gray!20}
Detector pixel & $P_d$ & $G_0$-$G_1$ distance& $G_1$-Det. distance & Step Res. & $D_{OD}$ & $p_0$ \\
\rowcolor{gray!20}
($\mu$m) & ($\mu$m) & $l$ (cm) & $d$ (cm)& ($\mu$m) & (cm) & ($\mu$m) \\
\hline
55  & 16 & 40 & 125 & 5  & 54 & 5.1\\
\hline
75  & 22 & 37 &  171 & 5  & 73 & 4.7 \\
\hline
150 & 44 & 33 & 343 & 10 & 147 & 4.2 \\
\hline
\end{tabular}
\label{tab:det cases parameters}
\end{table}

\textit{Indirect Detectors:}
Indirect scintillator-based detectors introduce spatial blur that can reduce fringe modulation and wash out high-frequency fringes. Therefore, the projected fringe period at the detector, $P_d$, can become a limiting factor for a given detector pixel size; for example, a 50$~\mu$m pixel may require a relatively large projected fringe period, such as $P_d > 130~\mu$m, to preserve measurable visibility. For Gaussian blur, the visibility reduction can be approximated as
$V = \exp\left[-\left(1.887\frac{w}{p_2}\right)^2\right]$, where $w$ is the FWHM of the point-spread function and $p_2=P_d$ is the fringe period at the detector \cite{bib:Weitkamp2006tomography}. This work, therefore, did not model the indirect detectors. 

\section{Image Recovery}
\label{subsec:image_recovery}

The image recovery algorithm is broken into two stages- \textit{interlacing the phase steps} and \textit{reconstruction} as shown in Figures \ref{fig:Rasterization process} and \ref{fig:Stages1and2}.  This combined technique is referred to as super-resolution, which leads us to perform analyzer-less X-ray interferometry.

\subsection{Interlacing}
\label{subsubsec: sub-pixel sampling}

\begin{figure}[h!]
    \centering
    \includegraphics[keepaspectratio=true, width=\textwidth]{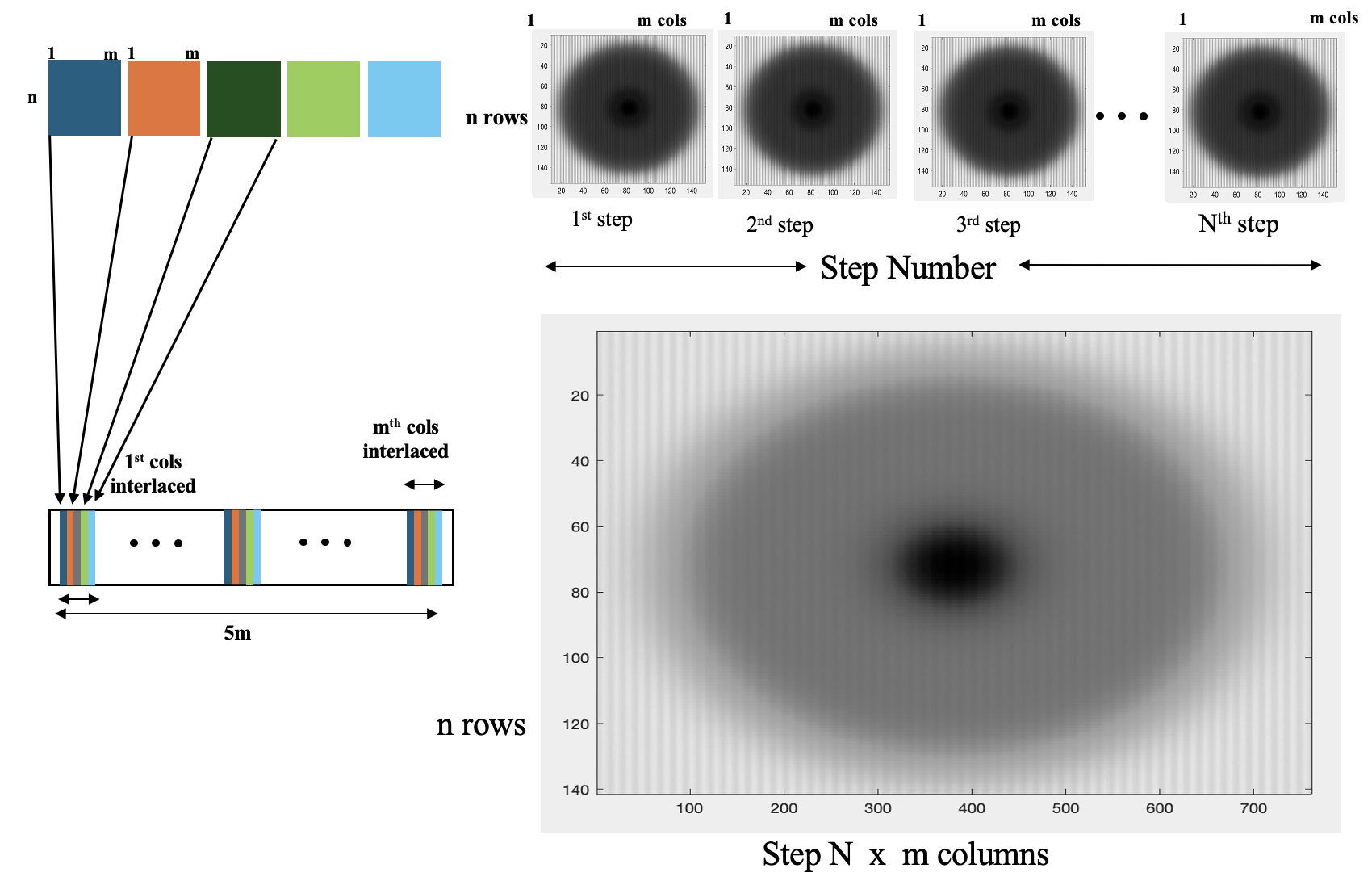}
    \caption{Left: Schematic representation of interlacing five phase-stepped images, $D^k_{obj}(x,y)$, k = 1...5, each with dimension n × m. The first columns from all five images are concatenated in order, followed by the second columns, and so on, until the last or m-th column. The resulting image, $I_{obj}(x,y)$, has dimensions n × 5m.
    Note: Although the phase-stepped images are shown in distinct colors for illustrative purposes, they are nearly identical in intensity in practice(representing either a set of reference phase steps or reference+object phase steps). The observed intensity variations arise primarily from sub-pixel phase shifts on the order of a few microns.
    Right: The figure shows the same process for the down-sampled interferometry signal. The downsampled signal, $D^k_{obj}(x,y)$, is interlaced to generate the interleaved image, $I_{obj}(x,y)$}

\label{fig:Rasterization process}
\end{figure}

The acquired detector signals or phase steps $D^k_{obj}(x,y)$ (each of size, say n x m) are interlaced in the x-direction to obtain an object image with nominal sub-pixel sampling of size n x Nm. This process is shown in Figure \ref{fig:Rasterization process}. Specifically, the first columns from phase steps k=1 to N are sequentially appended, followed by the second columns from each phase step, and so on, until all the columns have been assembled. This results in a larger n x Nm image,  $I_{obj}(x,y)$.  The same procedure is applied to the reference phase steps $D^k_{r}(x,y)$, producing a similarly interlaced $I_{r}(x,y)$.

At this stage, the interlaced images, $I_{r}(x,y)$ and $I_{obj}(x,y)$, are nominally sampled in the x-direction at the phase stepping resolution. The term nominal is used because even though the resolution is now higher and aliasing is nominally mitigated, the pixel blur remains, substantially reducing fringe visibility. This makes it difficult or impossible to recover the differential-phase and dark-field images using conventional algorithms at this stage.  To address this, we apply an iterative reconstruction algorithm to recover the visibility from the reference image $I_{r}(x,y)$, and use the visibility-restored reference in turn iteratively to recover the attenuation, differential-phase, and dark-field images from $I_{obj}(x,y)$, as detailed next.

\subsection{Iterative Reconstruction}

\begin{figure}[h]
    \centering
    \captionsetup[subfigure]{skip=5pt} 
    \begin{subfigure}[t]{0.42\textwidth}
        \centering
        \includegraphics[keepaspectratio=true, width=\textwidth]{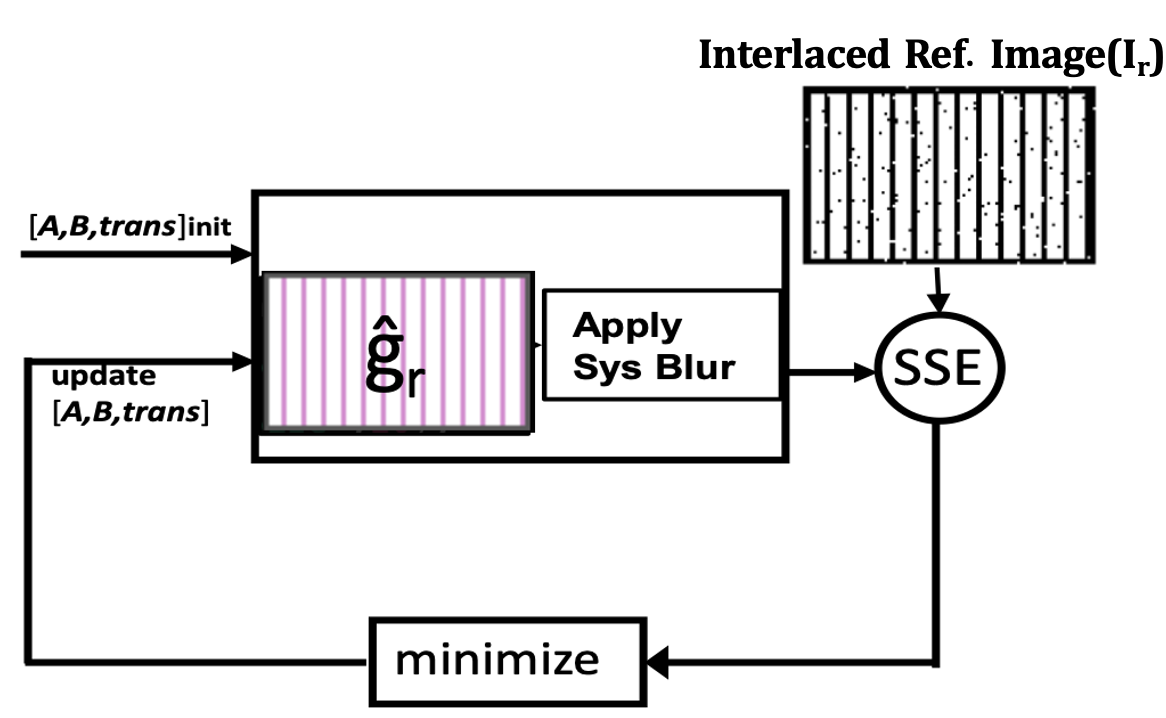}
        \subcaption{}
        \label{fig:Stage1}
    \end{subfigure}
    \hfill
    \begin{subfigure}[t]{0.54\textwidth}
        \centering
        \includegraphics[keepaspectratio=true, width=\textwidth]{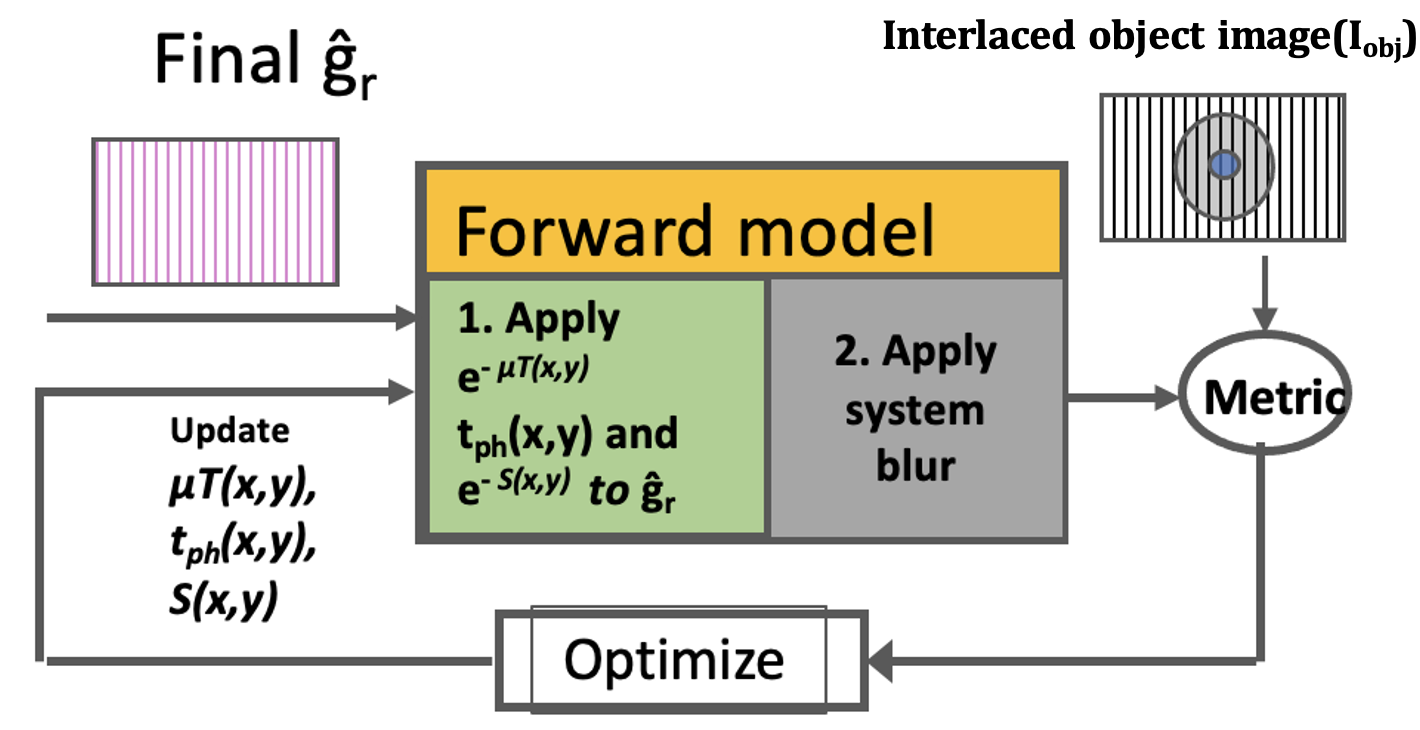}
        \subcaption{}
        \label{fig:Stage2}
    \end{subfigure}
    \caption{Stages 1 and 2 (a) Stage 1 to recover $\hat g_r$. This essentially recovers the visibility for the reference signal, $I_r(x)$.   (b) Stage 2 to recover $\mu T (x)$, $t_{ph}(x)$, and $S(x)$. }
    \label{fig:Stages1and2} 
\end{figure}

Reconstruction of the images was performed in two stages as shown in Figure \ref{fig:Stages1and2} using the resultant interlaced images, $I_r(x,y)$ and $I_{obj}(x,y)$. 

The first stage is the removal of the detector blur from the noisy reference image, $I_r(x,y)$, illustrated in Figure \ref{fig:Stage1}. Direct deconvolution produced oscillations and/or introduced apodization errors, so we performed this step iteratively, as is commonly done in super-resolution literature \cite{bib:SuperResFarsiu}.
The goal is to recover a visibility-restored estimate of the reference image, $\hat{g}_r(x,y)$, that has no detector blur.  This was done by iteratively fitting the initial parameters, [$\hat{A}(x,y), ~\hat{B}(x,y), ~\hat{\phi}_0(x,y)$], where $\hat{\phi}_0$ is the phase of the initial reference pattern.  While $\phi_0(x,y)$ is not needed in the image simulation, real experimental data would have an arbitrary phase, so $\hat{\phi}_0(x,y)$ must be included.  Starting with an initial estimate, $\hat{g}_r(x,y)$ is calculated and detector blur is applied.  The sum squared error (SSE) is calculated and minimized by iteratively updating [$\hat{A}(x,y), ~\hat{B}(x,y), ~\hat{\phi}_0(x,y)$] using \textit{fmincon} in Matlab until convergence is achieved \cite{bib:matlab_optim_toolbox}.

The second stage, illustrated in Figure \ref{fig:Stage2}, is the recovery of $\mu T(x,y)$, $t_{ph}(x,y)$, and $S(x,y)$ by iterative reconstruction, using the visibility-restored reference from Stage 1 in the forward model. Starting with an initial estimate of $\mu T(x,y)$, $S(x,y)$ and $t_{ph}(x,y)$, the simple forward model $\hat{g}_{obj}(x,y)$ is used to calculate the expected fringe pattern (that contains no pixel blur), shown in Eqn. \ref{eq:forward_model}. To limit the number of parameters of recovery, a simplified model is adopted, instead of the full spectral model in Eqn. \ref{eq:object_signal}.

Note the parameters [$\hat{A}, \hat{B}, \hat{\phi}_0$] finalized visibility-restored reference parameter from Stage 1, and the object parameters are used to construct the $\hat{g}_{obj}(x,y)$. To be able to compare with the real image, ${I}_{obj}(x,y)$, the blur is then simulated by convolution, producing the expected measurement, $\hat{I}_{obj}(x,y)$. The $\hat{I}_{obj}(x,y)$ and ${I}_{obj}(x,y)$ are then compared using a given metric and the metric is optimized.
The $PSF_d$ used to recover in Eqn. \ref{eq:blurred_forward_model} was a simple box blur while the generated $PSF_d$ contained the system degradation effects. 

The form of Eqn. \ref{eq:object_signal} suggests that $\mu T(x,y)$ can be estimated first, followed by $t_{ph}(x,y)$ and $S(x,y)$. 
Moreover, because $\mu T(x,y)$ attenuates both the bias and the amplitude (and it can be estimated from the average or the bias), it is the least noisy component, whereas $S(x,y)$ affects only the amplitude and is therefore more prone to noise.
 
The model parameters were estimated using multiple optimization strategies (adaptive gradient descent, Newton's method, quasi-newton method), and two metrics - the Maximum likelihood function (ML) and the Huber loss. 
Several optimization strategies were tested during development. Although a single objective function and optimizer can be applied to recover all three parameters, parameter-specific objective and optimizer choices were used to improve convergence stability.
The final implementation used Newton updates for ($\mu T(x,y)$), adaptive gradient descent for ($t_{ph}(x,y)$), and quasi-Newton minimization for ($S(x,y)$)\cite{bib:Broyden_Quasi_Newton}, with the Poisson maximum-likelihood objective used for ($\mu T(x,y)$) and ($t_{ph}(x,y)$), and the Huber loss used for ($S(x,y)$), because these choices provided stable convergence for the simulated cases considered here. Analytical derivatives were used in the implementation. Since the derivative involves the oscillatory intensity, it is smoothed via convolution by a window of width $2 P_d$ to $4 P_d$ so that the parameter estimates do not oscillate in x.

\begin{gather}
    \label{eq:forward_model}
    \hat{g}_{obj}(x,y) = \left\{ \hat{A}(x,y) e^{-S(x,y)} \cos \left( \frac{2\pi}{P_d} \left(x + t_{ph}(x) + \hat{\phi}_0 \right) \right) + \hat{B}(x,y)\right\} e^{-\mu T(x,y)} \\
    \label{eq:blurred_forward_model}
    \hat{I}_{obj}(x,y) = \hat{g}_{obj}(x,y) \star \star PSF_d(x,y) \star \star PSF_s(x,y)
\end{gather}

\begin{equation}
    \label{eq:likelihood}
    L(x) = I_{obj}(x)*\log{ \left(\hat{I}_{obj}(x)\right)} - \hat{I}_{obj}(x)
\end{equation}

\begin{equation}
\label{eq:Huber Loss}
L_\delta(r) = 
\begin{cases} 
\dfrac{1}{2} r^2 & \text{if } |r| \leq \delta, \\[6pt]
\delta \left(|r| - \tfrac{1}{2}\delta\right) & \text{if } |r| > \delta
\end{cases}
,\quad
r = \hat{I}_{\text{obj}}(x) - I_{\text{obj}}(x).
\end{equation}

\section{Results}
\label{sec:results}

\begin{figure}[htbp]
    \centering
    \includegraphics[width=1\textwidth]{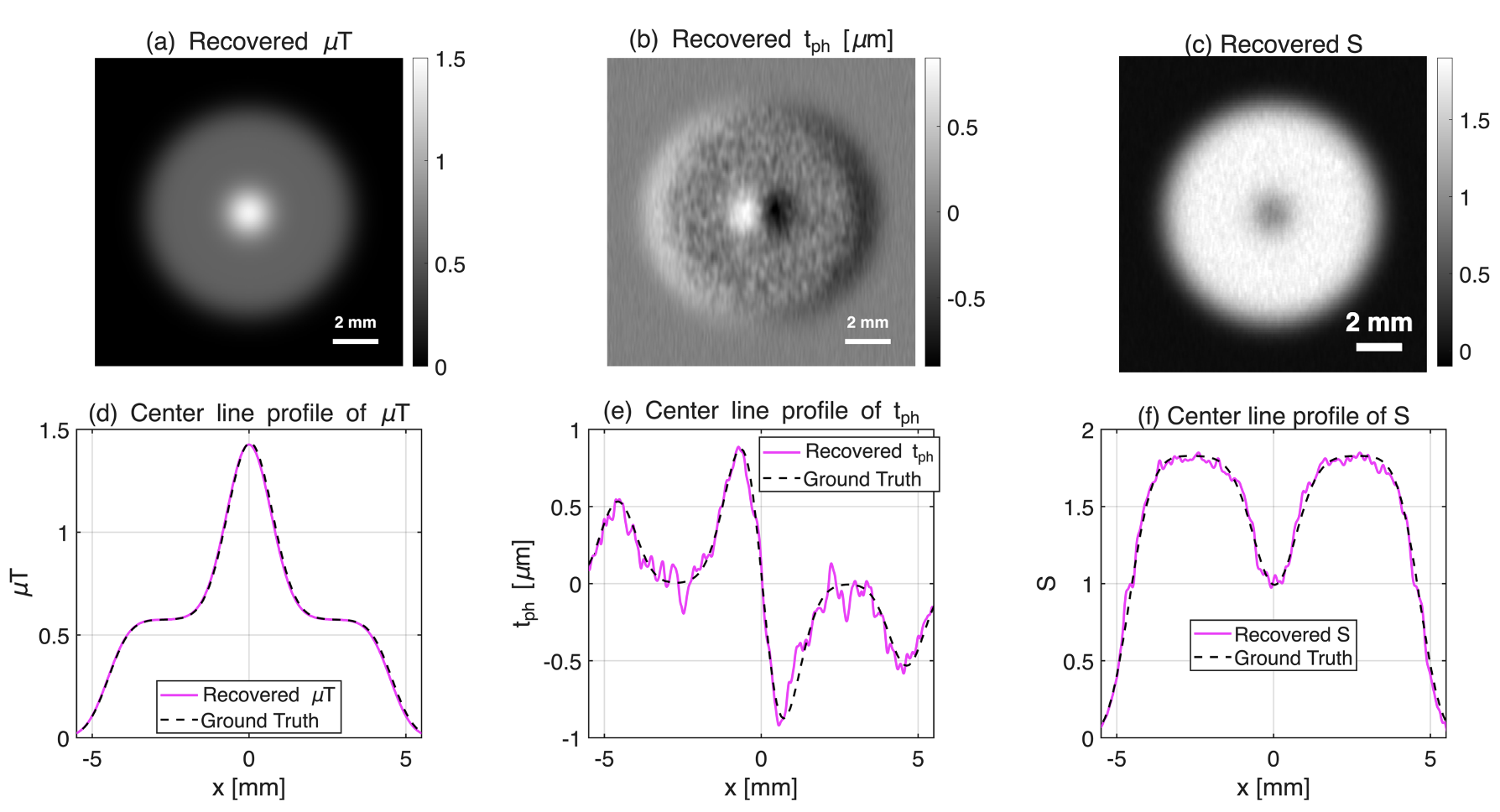}
    \caption{Recovered attenuation $\mu T$, phase-shift $t_{\mathrm{ph}}$, and dark-field $S$ maps are shown in the top row. The corresponding center line profiles are shown in the bottom row, where the recovered profiles are plotted in purple, and the ground-truth profiles are plotted as black dashed lines.}
    \label{fig:det75Pd22}
\end{figure}

The reconstruction for the case of detector pixel size $75 ~\mu m$ with a $P_d=22 ~\mu m$ is shown in Fig. \ref{fig:det75Pd22}. Attenuation, differential phase-shift and dark-field images are well recovered with good agreement to the ground truth. The bottom row of Fig. \ref{fig:det75Pd22} shows the line profile through the center of the image. 

\begin{figure}[htbp]
    \centering
    \includegraphics[width=1\textwidth]{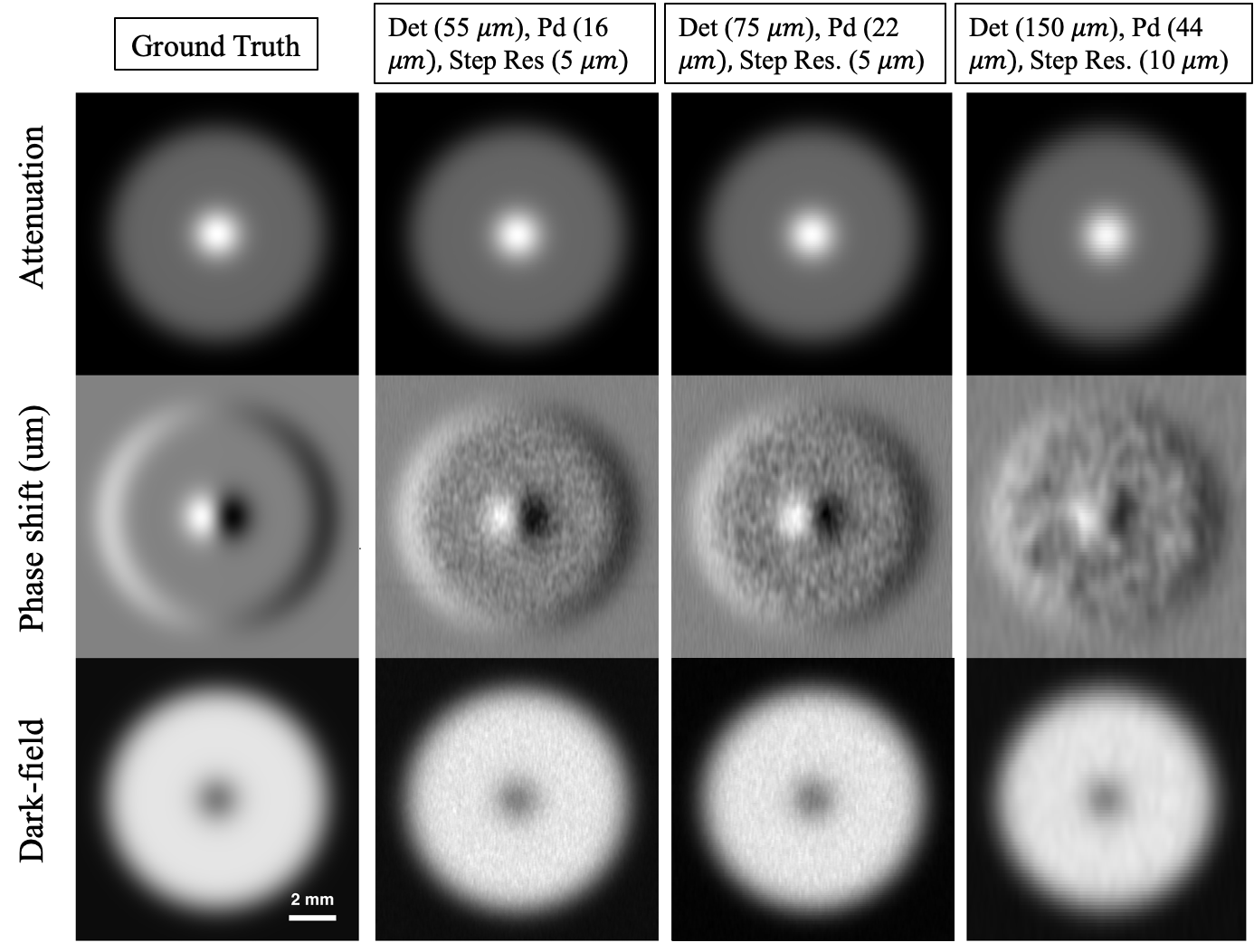}
    \caption{Image parameter recovery of various cases compared with the ground truth for the same Nyquist undersampling ratio (2$\times$ Det. size/$P_d$).}
    \label{fig:comparison_grid}
\end{figure}

The other cases of detector pixel sizes ranging from $55-150 ~\mu m$ with corresponding periods ($P_d$) are shown in Fig. \ref{fig:comparison_grid}. The $P_d$ selected are such that the Nyquist undersampling ratio (2$\times$Det.Pixel size/$P_d$) is the same. Even though the Nyquist ratios are the same, these cases still differ by source size magnification, step resolution and importantly, detector pixel size. Among the three recovered modalities, attenuation and dark-field showed the most stable recovery, while the phase-shift image was more sensitive to detector pixel size, noise, and residual model mismatch.

For the dark-field parameter, $S$, the autocorrelation length was kept approximately constant across detector cases by varying $D_{OD}$ with $P_d$, allowing the same dark-field coefficient model to be used for comparison. Because ($t_{ph}(x,y) = \frac{\lambda}{2\pi}\frac{\partial \varphi(x,y)}{\partial x}D_{OD}$), this geometry change would also alter the differential-phase-shift magnitude. In the present simulations, however, the phase-shift map was held fixed to isolate the effect of detector sampling and reconstruction. The corresponding maximum phase shift would otherwise range from approximately (0.63 to 3.3~$\mu m$), while the plotted values represent the mean recovery of the fixed phase-shift ground truth.

\begin{figure}[h!]
    \centering
    \includegraphics[keepaspectratio=true, width=0.5\textwidth]{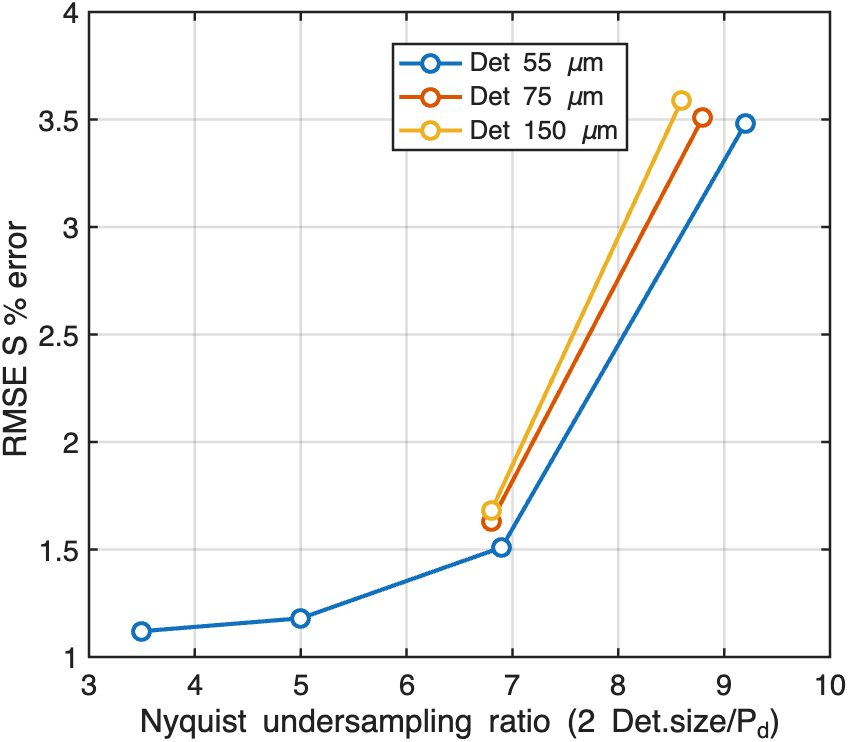} 
    \caption{Normalized RMSE error in the recovered dark-field parameter $S$ as a function of the Nyquist undersampling ratio, for the detector cases considered.}
    \label{fig:error plot}
\end{figure}

The RMSE of the recovered dark-field parameter, $S(x,y)$, is plotted in Fig.~\ref{fig:error plot} as a function of the Nyquist undersampling ratio, DetPixel/($P_d$/2), or 2DetPixel/$P_d$. The error in $S(x,y)$ increases as the Nyquist ratio increases, indicating that dark-field recovery becomes more difficult as the projected fringe period becomes smaller relative to the detector pixel size. Therefore, for a given detector pixel size and reconstruction condition, there is a lower practical limit on $P_d$ below which the super-resolution recovery becomes unreliable.

The error metrics for the images shown in Fig.~\ref{fig:comparison_grid} are summarized in Table~\ref{tab:error table}. The normalized root-mean-square error (NRMSE) and normalized mean absolute error (NMAE) were calculated over the analysis ROI, normalized by the dynamic range of the ground-truth image, and reported as percentages. MSSIM denotes the mean structural similarity index between the recovered and ground-truth images.

\begin{figure}[htbp]
    \centering
    \includegraphics[width=1\textwidth]{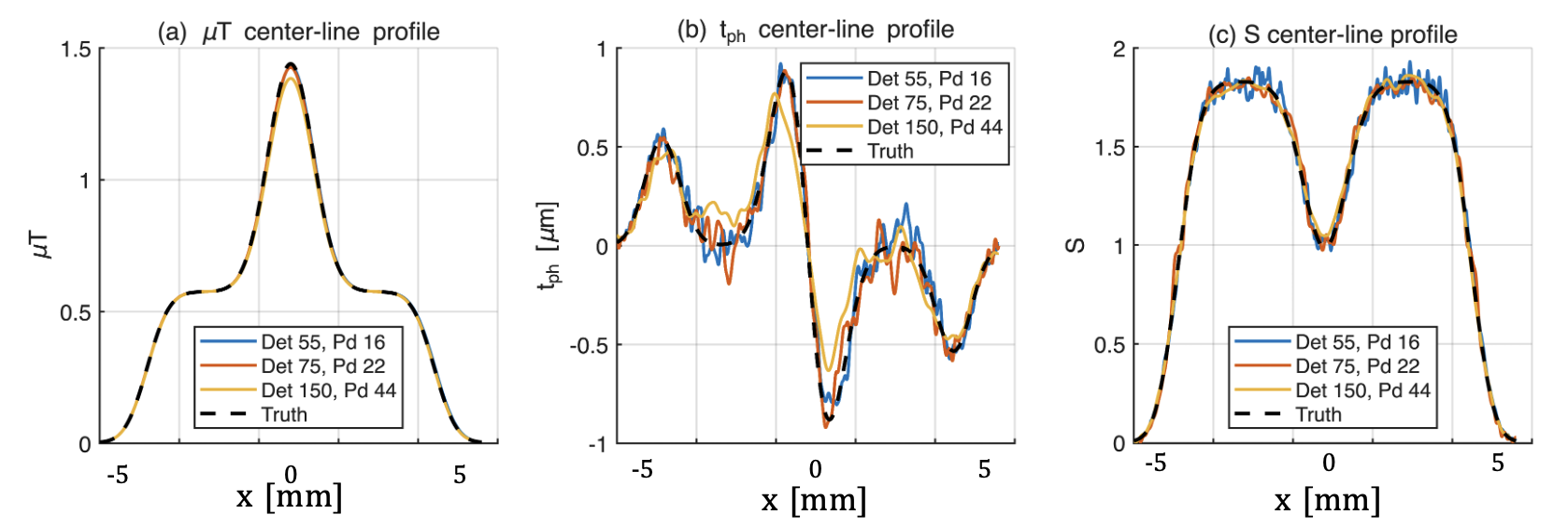}
    \caption{Line profiles of the four cases shown in Fig. \ref{fig:comparison_grid}.}
    \label{fig:lineProfiles}
\end{figure}

\begin{table}[htbp]
\centering
\caption{Comparison of reconstruction error metrics for different detector cases.}
\label{tab:error_metrics}

\renewcommand{\arraystretch}{1.25}
\setlength{\tabcolsep}{7pt}

\begin{tabular}{|l|ccc|ccc|ccc|}
\hline
\cellcolor{gray!35} 
& \multicolumn{3}{l|}{\cellcolor{gray!35}\textbf{NRMSE in \%}} 
& \multicolumn{3}{l|}{\cellcolor{gray!35}\textbf{NMAE in \%}} 
& \multicolumn{3}{l|}{\cellcolor{gray!35}\textbf{MSSIM}} \\
\cline{2-10}

\cellcolor{gray!35}
& \cellcolor{gray!35}\textbf{\textit{$\mu T$}}
& \cellcolor{gray!35}\textbf{\textit{$t_{ph}$}}
& \cellcolor{gray!35}\textbf{\textit{$S$}}
& \cellcolor{gray!35}\textbf{\textit{$\mu T$}}
& \cellcolor{gray!35}\textbf{\textit{$t_{ph}$}}
& \cellcolor{gray!35}\textbf{\textit{$S$}}
& \cellcolor{gray!35}\textbf{\textit{$\mu T$}}
& \cellcolor{gray!35}\textbf{\textit{$t_{ph}$}}
& \cellcolor{gray!35}\textbf{\textit{$S$}} \\
\hline
Det 55 $\mu$m, $P_d$ 16 $\mu$m & 0.582 & 2.49 & 1.51 & 0.325 & 1.71 & 1.12 & 0.998 & 0.715 & 0.929 \\
\hline
Det 75 $\mu$m, $P_d$ 22 $\mu$m & 0.405 & 3.08 & 1.63 & 0.218 & 2.14 & 1.18 & 0.998 & 0.646 & 0.916 \\
\hline
Det 150 $\mu$m, $P_d$ 44 $\mu$m & 0.909 & 4.27 & 1.68 & 0.491 & 2.98 & 1.22 & 0.991 & 0.585 & 0.910 \\
\hline
\end{tabular}
\label{tab:error table}
\end{table}
\section{Discussion}

\label{sec:discussion}

We demonstrated in simulations that super-resolution iterative reconstruction can recover radiographic attenuation, differential-phase, and dark-field images using data with different detector resolutions and Nyquist undersampling ratios for analyzer-less grating systems. The quantitative comparison showed promising results: all three modalities are recoverable without the analyzer grating used in conventional Talbot--Lau X-ray interferometry, with an error of up to 4\%. Among the three recovered modalities, attenuation and dark-field showed the most stable recovery, while the phase-shift image was more sensitive to detector pixel size, noise, and residual model mismatch. 

Data were generated using the full spectral model (Eqns. \ref{eq:object_signal} and \ref{eq:blurred_object_signal}), whereas recovery was performed using a simplified monochromatic model (Eqns. \ref{eq:forward_model} and \ref{eq:blurred_forward_model}) to limit the number of estimated parameters. Thus, the observed errors may partly reflect inaccuracies introduced by this modeling approximation.

Phase-stepping errors from detector stepping are unavoidable in practice. To evaluate their effect, a 20 $\%$ stepping uncertainty ($\pm 1~\mu$m for a nominal 5$~\mu$m step) was tested. The imaging parameters were still successfully recovered, although small Moiré artifacts were observed(not shown). Artifact mitigation strategies previously investigated by our group for TLI \cite{Meyer2025moire}, including multi-harmonic corrections and regularization, could be incorporated into the present iterative reconstruction framework for both reference and object images.

A limitation of this method is that larger detector pixels require sufficiently large projected fringe periods to keep the Nyquist undersampling ratio within a recoverable range. This limits $P_d$ for a given detector and, therefore, the autocorrelation length of the system. One limitation of the model is that it does not perform a full ray-traced simulation of the object, including Compton scattering. In future work, this could be incorporated into the analytical with-object fringe model. Compton scatter may be modeled as a bias term. In the recovery, it can be removed algorithmically \cite{Smith2025MLEGF}. 

As shown in Fig.~\ref{fig:error plot}, the practical lower limit on ($P_d$) can be inferred from the increase in recovery error of $S$ with increasing Nyquist undersampling ratio. This geometric constraint is most restrictive for larger-pixel detectors. This fringe-magnification requirement can constrain the system geometry and experimental setup. In addition, the analyzer-less super-resolution approach is primarily suited to direct detectors; implementing it with indirect scintillator-based detectors would require substantially larger projected fringe periods, typically $P_d > 100~\mu$m, to preserve measurable fringe visibility in clinically relevant detector systems.

The super-resolution method will benefit other interferometry systems. For example, the Modulated Phase Grating system \cite{bib:MeyerDeySciRep2024}, which has an analyzer-less design, will benefit from this method by enabling a smaller Pd (i.e., higher ACL) for a given detector pixel size. Future work will include extending the analysis to anatomically realistic phantoms, three-dimensional interferometric computed tomography for Modulated Phase Grating and Talbot Lau systems, and ultimately experimental verification and clinical adoption.

\section{Conclusion and Future Work}
\label{sec:conclusion}

Our simulations demonstrate that super-resolution can resolve attenuation, differential-phase, and dark-field images with grating interferometers even when detector sampling falls below the Nyquist rate required by traditional image recovery algorithms.  Iterative reconstruction was shown to recover images of a two-dimensional lung phantom with a diseased region from simulated data. In the Talbot–Lau Interferometer, this offers the potential for dose reduction for a fixed image quality, while simplifying system alignment and reducing cost by eliminating the analyzer grating. The method presented extends applicability to cases where traditional algorithms fail. 
In the future, we will focus on applying this method to other analyzer-less interferometry systems, such as Modulated Phase Grating (MPG) interferometry, as well as the clinical translation of these multi-modal X-ray interferometry systems.
\section{Acknowledgments}

This work is funded in part by NIH NIBIB Trail-blazer Award 1-R21-EB029026-01A1. An earlier version of this work was presented at the 67th Annual Meeting \& Exhibition of the American Association of Physicists in Medicine (AAPM 2025) \cite{Taqi2025AAPM}.

\section{Author Contributions} 
  MST constructed the two-dimensional phantom, the interferometric data generation and reconstruction algorithms, implemented the 2D iterative reconstructions, and performed the data analysis. HCM incorporated the differential-phase parameter and contributed to the development of the initial framework and phantom. JD conceived the method, developed the initial framework, guided the development of the algorithm, and supervised the research. All authors contributed to the writing, discussion, and revision of the manuscript.
\section{Conflict of Interest}
JD is an inventor of two patents related to the Modulated Phase Grating Interferometry \cite{bib:MPGPatent1,bib:MPGPatent2}.

\newpage
\printbibliography

\end{document}